\documentclass[12pt]{article}

\usepackage{epsfig}
\usepackage{graphicx}

\title{Core-halo instability in dynamical systems} 

\author
{Seth Lloyd\\
\\
\normalsize{Department of Mechanical Engineering}\\
\normalsize{MIT 3-160, Cambridge MA 02139 USA}\\
\normalsize{The Santa Fe Institute}\\
\normalsize{1399 Hyde Park Road, Santa Fe, NM 87501 USA}\\
\\
\normalsize{slloyd@mit.edu}
}

\date{}

\begin{document}


\baselineskip24pt


\maketitle

\begin{abstract}
\noindent This paper proves a set of instability theorems 
for dynamical systems. 
As interactions are added between subystems in a complex
system, structured or random,
a threshold of connectivity is reached beyond which
the overall dynamics inevitably either becomes highly oscillatory,
unstable, or both.   The threshold
occurs at the point at which flows and interactions
between subsystems (`surface' effects) overwhelm
internal stabilizing dynamics (`volume' effects).
The theorems are used to identify  oscillagion/instability thresholds
in systems that possess a core-halo or core-periphery structure, including
the gravo-thermal catastrophe -- i.e., star collapse and explosion --
and the interbank payment network.   In the core-halo model,
the same dynamical instability underlies both gravitational and 
financial collapse.
\end{abstract}

A wide variety of work addresses the stability of dynamical systems 
made up of networks of interacting subsystems [1-5].  A key ingredient 
of stability is network connectivity [5].  One of the best-known
results in this field is May's theorem that 
differential equations described by
random networks undergo a transition from stable to unstable 
behavior at a critical value of their connectivity [4].  As
May himself noted [4], networks that occur in nature are rarely random:
they typically possess complex structures related to their function [5].
In non-random networks, for example, realistic models of food webs
[6], adding connections may either stabilize or destabilize the network.
This paper prove a sequence of instability theorems for dynamical systems
described by structured,
non-random networks and applies that theorem to dynamical systems
that possess a dense core surrounded by a diffuse halo
(the term used in astrophysics and
elementary particle physics [7-9]) or periphery (the term
used in economics and social sciences [10-13]).  Two such systems are
the interbank transfer network [10-11], and the network of gravitational
interactions within a star [7-8].  As interactions are added between
core and halo, the overall system either undergoes increasingly 
underdamped oscillations, or goes unstable, or both.

The stability to instability transition identified in this
paper arises because excess connectivity drives instability,
not just for differential equations with random gradients
as in [4], but for any set of
coupled ordinary differential equations.  
Such sets of equations are ubiquitous
in the mathematical modeling of dynamical systems, and can be
applied to physical systems (e.g., Newtonian gravity, electrodynamics),
networks of chemical reactions, biological systems (e.g., ecological models
and food webs), engineered systems (feedback control),
interacting agents [15],
and economic and financial systems (market
economies, flows of money and debt).  

\bigskip\noindent{\it Interactions and instability}

Consider a set of non-linear, time-dependent, ordinary differential equations
over $n$ variables:
$$ {d\vec x\over dt} = g( \vec x, t),\eqno(1)$$ 
where ${\vec x} = (x_1, \ldots, x_n)$.  
The dynamics of a small perturbation $\Delta \vec x(t)$ to
a solution $\vec x^*(t)$ obeys the linearized equation
$$\Delta \dot{\vec x} = \nabla g|_{\vec x^*}  \Delta \vec x.\eqno(2)$$
The perturbation decreases in size if
$$ {d\over dt}\Delta \vec x^\dagger \Delta \vec x 
= \Delta \vec x^\dagger ( \nabla g^\dagger + \nabla g) \Delta \vec x 
\equiv 2 \Delta \vec x^\dagger G \Delta \vec x <  0, \eqno(3)$$
where $G = ( \nabla g^\dagger + \nabla g) /2$
is the symmetrized Hermitian gradient evaluated at $x^*(t)$.
The Hermitian part of the gradient governs exponentially increasing and
decreasing behavior, while the anti-Hermitian part 
$\tilde G = ( \nabla g^\dagger - \nabla g)/2$ governs
oscillatory behavior.
All perturbations decrease in size if and only if the Hermitian
gradient $G$ is negative definite.    The threshold of
instability identified in this paper occurs at the point
where interactions (off-diagonal terms in $\nabla g$ and $G$)
become sufficiently strong to make some eigenvalue
of $G$  positive, so that some perturbations grow in size. 
Similarly, the dynamics becomes underdamped when interactions
become sufficiently strong that an oscillatory eigenvalue
of $\tilde G$ becomes larger than the damping rate of the
corresponding eigenstate. 
Note that the definition of stability adopted here -- small perturbations
decrease in size -- is stronger than Lyapunov's definition of
stability, which demands only that 
small perturbations {\it eventually} decrease in size [1-2]. 

The instability/oscillatory thresholds identified here arise from
excess of interaction.  Off-diagonal terms in $\nabla g$ govern 
interactions or flows of energy, entropy, money, etc., 
between subsystems of the network, and on-diagonal terms
represent sources and sinks of the same quantities.
The internal dynamics of subsystems correspond to
diagonal blocks of the gradient matrix of the linearized equations,
while flows between subsystems correspond to off-diagonal blocks.
Look at the interaction between two such subsystems.
Subsystem ${\cal A}$ consists of $n_A$ variables, 
and subsystem ${\cal B}$ consists of $n_B \geq n_A$ variables.
Let $G_{AB}$ be
the restriction of $G$ to the subspace spanned by the
$n = n_A + n_B$ variables that describe  ${\cal A}$, ${\cal B}$. 
Assume that ${\cal A}$ and ${\cal B}$ are locally stable in the
absence of interaction, and investigate how that stability
changes as interactions are added.  
The interactions between subsystems lead to stable
dynamics if all the eigenvalues of the Hermitian matrix $G_{AB}$ 
are negative.  Write this matrix as
$$ G_{AB} = \pmatrix{ A & C \cr C^\dagger & B \cr},\eqno(4)$$
where $A$ gives the Hermitian dynamics confined
to the subsystem ${\cal A}$, and $B$ gives the Hermitian dynamics
for ${\cal B}$.  By the assumption of local stability, 
$A$ and $B$ are negative definite.
$C$ is an $n_A \times n_B$ matrix whose 
coefficients determine the strength of interactions 
between ${\cal A}$ and ${\cal B}$.  

As more and more interactions are added, and as the strength
of those interactions increase, then the interactions inevitably
drive the system unstable.  In particular, we have

\bigskip\noindent{\it Theorem 1:}
If ${\rm tr}~ C^\dagger C > \sqrt{ {\rm tr} A^2}
\sqrt{ {\rm tr} B^2}$, 
then the system is unstable.

\bigskip\noindent 
Theorem 1 is a higher dimensional generalization of
the fact that a $2\times 2$  matrix 
$\pmatrix{\mu  & \lambda \cr \bar\lambda  & \nu \cr}$
where $\mu,\nu < 0$,  has a positive
eigenvalue if $|\lambda|^2 > \mu\nu$.  

Theorem 1 states that the interacting systems are unstable
when the average magnitude squared of the terms in the destabilizing
interactions is larger than the geometric mean
of the average magnitude squared of the terms in
the stabilizing local dynamics.
Inevitably, if the strength of the
stabilizing local dynamics is fixed, increasing
the strength of the interactions drives
the system unstable beyond some threshold.  
Intuitively, the threshold
occurs at the point where flows through the `surface' between ${\cal A}$ and
${\cal B}$ dominate the `volume' flows within ${\cal A}$ and ${\cal B}$. 
Applied to random matrices representing the interactions between
two parts of a complex system, theorem 1 reproduces the 
results of May [4] for connection-driven instability.  
However, no assumptions concerning
random matrix theory were required to prove the 
theorem -- the matrices involved can be highly structured.  
For example, in food webs [6], the hierarchical nature of who eats
whom leads to structured networks: adding additional species
and connections can increase the magnitude of both the off-diagonal terms of
the gradient and of the on-diagonal terms, potentially leading
to greater stability rather than instability.     

Theorem 1 is a sufficient condition for instability.   In many systems
the onset of instability could occur at a much lower threshold for the
strength of the off-diagonal terms.   We can make the bounds of 
theorem 1 tight as follows:   

\smallskip\noindent (1a) If ${\rm tr} C^\dagger C < \min \mu_i \nu_j$,
the minimum product of eigenvalues $\mu_i$, $\nu_j$ of $A$, $B$,
then the system is stable.

\smallskip\noindent (1b) For fixed
${\rm tr} C^\dagger C >  \min \mu_i \nu_j$,
the coefficients of $C$ can always be chosen to make the system unstable.

\smallskip\noindent (1c) For fixed  $ \sqrt{ {\rm tr} A^2}
\sqrt{ {\rm tr} B^2} >  {\rm tr} C^\dagger C$, the coefficients of
$A,B$ can always be chosen to make the system stable.

\bigskip
We now prove a similar threshold theorem for oscillatory behavior.
Write the anti-Hermitian gradient as
$$\tilde G_{AB} = 
\pmatrix{ \tilde A & \tilde C \cr  -\tilde C^\dagger & \tilde B \cr}.\eqno(5)$$
Define the degree of underdamping $\gamma$ for
at state $\vec w$ 
to be equal to the ratio between the oscillatory
rate and the damping rate,
$|\vec w^\dagger \tilde G_{AB}\vec w|
/ |\vec w^\dagger G_{AB} \vec w|$ (assuming
that the state is stable so that $\vec w^\dagger G_{AB} \vec w < 0$).
The degree of underdamping for the system is the maximum degree
of underdamping over all states $\vec w$.
Our second theorem shows that as more and more terms are added to
the off-diagonal part of the
anti-Hermitian gradient, $\tilde C$, the system becomes 
increasingly underdamped:

\bigskip\noindent{\it Theorem 2:}
If $ ~{\rm tr}~ \tilde C^\dagger \tilde C \geq \gamma^2 \sqrt{ {\rm tr} A^2}
\sqrt{ {\rm tr} B^2}$,
then the system has underdamping degree at least $\gamma$.

\bigskip\noindent
Theorem 2 also yields tight bounds for oscillatory behavior in analogue
to (1a) (1b) (1c).
Since interaction terms contribute either to Hermitian gradient
interaction ${\rm tr}  C^\dagger C$ or to the anti-Hermitian gradient 
${\rm tr} \tilde C^\dagger \tilde C$ or to both, theorems 1 and 2
imply that adding interactions
between ${\cal A}$ and ${\cal B}$ leads either to increasingly undamped
oscillations or to instability or to both.

Theorems 1 and 2 apply to any two subystems of a larger dynamical system.
They will now be used to analyze
interaction-induced oscillation/instability thresholds in systems
where ${\cal A}$ corresponds to a dense, highly interacting core, 
and ${\cal B}$ corresponds to a diffuse, weakly interacting halo.



\bigskip\noindent{\it Core-halo instability and
the gravo-thermal catastrophe}

A common type of system in the universe consists of a collection of
matter, e.g. a cloud of interstellar dust, a star, or a cluster of
stars in a galaxy, interacting
via the gravitational force, augmented by collisions and heat
production via, e.g., nuclear reactions.  Such a system
naturally forms itself into a dense `core' (system ${\cal A}$)
of strongly interacting matter at high temperature, surrounded  
by a less-dense `halo' (system ${\cal B}$).  The microscopic dynamics of
such a system are complex [7-8].    
A simple linearized model of the 
energy transfer dynamics between ${\cal A}$ and ${\cal B}$ 
in terms of macroscopic variables  
takes the matrix form (see supplementary material):
$$ {d\over dt}\pmatrix{ T_A\cr T_B\cr}
= \pmatrix{ (\eta - \alpha)/C_A &\alpha/C_A \cr 
\alpha/C_B & (-\zeta -\alpha)/C_B \cr}  \pmatrix{ T_A\cr T_B\cr} .\eqno(6)$$
Here, $T_A$ is the temperature of the core
and $C_A$ is its specific heat.
Similarly, $T_B$ and $C_B$ are the temperature and
specific heat of the halo.  $\alpha \geq 0$ gives the linearized rate of energy
transfer between core and halo as a function of their temperature difference.  
$\eta \geq 0$ governs energy production in the core, due, e.g., to nuclear
reactions, and $\zeta \geq 0$ governs heat loss from the halo to space
beyond.  

The key feature of equation (6) is that the specific heat of 
systems whose dynamics is dominated by gravity
is typically {\it negative}: when the hot core of tightly bound
particles loses energy, the remaining particles cluster
together more tightly and move {\it faster}.  When $C_A < 0$, demanding
that the system be locally stable and below the interaction-driven
instability threshold requires $C_B > 0$ and $\eta > \alpha$.
That is, the overall system
can still be stable if the specific heat of the halo is positive, 
so that like an ordinary gas it grows cooler as it loses energy, 
and if internal heat production in the core
outweighs heat loss to the halo.   As the 
internal rate of heat production slows -- for example,
as the nuclear reactions inside a star burn through their
fuel -- the system goes unstable at the critical threshold when $\eta$
becomes less than $\alpha$.  At this point destabilizing flows of energy from
core to halo dominate the stabilizing production of energy
within the core. 
The temperature of the core now rises exponentially in time, with 
exponentially increasing flows of energy from core to halo.  

Note that when $C_A < 0$, the anti-Hermitian part of the gradient
in equation (6) is larger in magnitude than the Hermitian part.
As a result of theorem 2, then,
before reaching the instability threshold, the core-halo
system will undergo increasingly underdamped oscillations.

The onset of oscillations leading to an accelerating, unstable flow of energy
is called the gravo-thermal catastrophe [7]: from the dynamics
(6) the gravo-thermal catastrophe is seen to be a straightforward
instance of interaction-driven oscillation and
instability governed by theorems 1 and 2.  
For a star with more than a few solar masses or for galaxy formation 
in the early universe, the gravo-thermal catastrophe 
results in gravitational collapse of the core, 
and the formation of a black hole.   With the formation of a 
black hole, energy flows from core to halo cease (except for
a small amount of Hawking radiation).  The black hole `freezes'
the previously hot core, and reverses the direction of energy flow,
sucking up matter and energy from the halo.

\bigskip\noindent{\it Core-halo instability in financial collapse}

Like galaxies or nebulae, the interbank payment transfer network possesses
a core-halo structure [10-11], and is susceptible to 
interaction-driven oscillation and instability.   
As detailed in [10], in 2007
this network consisted of over $6600$ 
financial institutions connected 
by over $70,000$ daily transfers.  Most of the institutions
(the halo) had either few links or links whose transfers had only
small volume.  A small,
highly connected fraction of the institutions (the core), accounted
for most of the volume.  On a typical day, for example,
a core of $66$ institutions connected by $181$ links comprised $75\%$ 
of the value transferred [10].  The core itself contained an `inner' core
of $25$ institutions that were almost fully connected.  
The core-halo structure of the network is shown in figure (1).
Most important for stability analysis, the interbank payment
transfer network is strongly disassortative [14]: 
highly-connected banks do most of their business with 
sparsely-connected banks, and {\it vice versa}. 
The disassortative nature
of the network means that there are fewer internal links
within the core, and within the halo,
than there are between core and halo.

Disassortative networks are known to be less stable
than assortative networks with respect to mixing and link removal [14].
The results of this paper can be applied to disassortative
networks corresponding to the dynamics of coupled ordinary differential
equations in general, and to the interbank transfer network
in particular.  Define a weighted disassortative network
to be one in which the weighted sum of the links between
highly connected core and the sparsely-connected halo is
greater than the weighted sums of
the links within the core and within the halo:
$\sum_{i\in C,j\in H} |c_{ij}|^2 >
\sum_{ij\in C} |a_{ij}|^2 + \sum_{ij\in H} |b_{ij}|^2$.
Theorems 1 and 2 then imply

\bigskip\noindent{\it Theorem 3:}  A dynamical system
whose gradient corresponds to a weighted disassortative network
is either underdamped (with $\gamma > 1$) or unstable.

\bigskip\noindent
Since the overall interbank network is disassortative,
theorem 3 implies that stability/lack of oscillation can only be obtained 
when the banks in the core have significantly stronger interactions
with each other than with banks in the halo.  That is,
even though the banks in the core undergo more transactions
with banks in the halo than with each other, to maintain stability, 
typical flows between banks in the core and 
and other banks in the core must be significantly larger than
typical flows between banks in the core and banks in the halo.
This `hot core' requirement for stability is confirmed by the data [10] --
as noted, the core contains three quarters of the flow on a given day.
Only by having large exchanges with each other (e.g., by hedging)
can the banks within the core overcome the disassortative
nature of the network to provide stability.  

The hot core requirement 
leaves the network vulnerable to interaction-driven
instability.  Theorems 1 and 2 imply that if some event causes a sudden drop in
the strength of transfer rates within the core, then 
the whole system can go unstable.  The mathematical origin 
of this financial instability is the same as the origin of instability
in gravo-thermal collapse, where a slowing of energy production in
the core drives the system unstable. The end point of the financial
instability is the well-known
liquidity trap, a spectacular example of which occurred during
the financial crisis of 2008-2009.  
For the bank transfer network, 
just as for black hole formation, instability 
leads to a collapsed regime in which the core freezes up, and 
transfers drop dramatically (`the black hole of finance').

\bigskip\noindent{\it Conclusion} 

This paper presented simple
mathematical criteria for the stability and oscillatory
behavior of dynamical systems 
as interactions are added between subsystems.  If the number and strength of 
interactions between subsystems grows too large, the criteria
identifies a threshold of connectivity beyond which oscillations
and/or fluctuations inevitably grow.
This result extends the May theorem for random networks
to structured networks.   A number of artificial and
naturally occurring dynamical systems, such as the gravitational
and financial systems discussed here, are subject to this threshold. 
Indeed, any system within which the number and strength of 
interactions increase over time, without an attending
increase in the strength of local stabilizing dynamics, will 
inevitably approach the interaction oscillation/instability threshold.
An interaction-driven core-halo instability lies at the 
heart of both financial and gravitational collapse.  
Once the instability threshold is passed, 
unless interactions between core and halo are reduced,
and local stabilizing dynamics within the core and halo are increased,
rapidly growing oscillations/fluctuations will 
overwhelm physical and financial stabilization mechanisms leading
to gravitational and financial collapse.

\vfill
\noindent{\it Acknowledgments:} This work was supported by a Miller
fellowship from the Santa Fe Institute.  The author would like to
thank Olaf Dreyer, Jeffrey Epstein, Doyne Farmer, Thomas Lloyd, Cormac McCarthy,
Sanjoy Mitter, Chris Moore, Sam Shepard, and Jean-Jacques Slotine for helpful 
conversations.
\vfil\eject

\noindent{\it References}

\bigskip\noindent
[1] D.G. Luenberger,
{\it Introduction to Dynamic Systems: Theory, Models and Applications,}
Wiley, New York, (1979).

\bigskip\noindent
[2] J.-J.E. Slotine, W. Li, {\it Applied Nonlinear Control},
Prentice Hall, Englewood Cliffs (1991).

\bigskip\noindent
[3] S. Strogatz, {\it Nonlinear Dynamics and Chaos}, Perseus Books,
Cambridge (1994).

\bigskip\noindent
[4] R. May, {\it Nature} {\bf 238}, 413 (1972).

\bigskip\noindent 
[5] M. Newman, D. Watts, A.-L. Barab\'asi,
{\it The Structure and Dynamics of Networks}, 
(Princeton University Press, 2006). 

\bigskip\noindent
[6] D. Angelis, {\it Ecology} {\bf 56}, 238-243 (1975).





\bigskip\noindent [7] D. Lynden-Bell, R. Wood, 
{\it Mon. Not. R. Astr. Soc.} {\bf 138}, 495-525 (1968).

\bigskip\noindent [8] M.P. Leubner, {\it Astrophys. J.} {\bf 604}, 469 (2004).

\bigskip\noindent [9] S. Nickerson, T. Csorgo, D. Kiang,
{\it Phys. Rev. C} {\bf 57}, 3251-3262 (1998); arXiv:nucl-th/9712059.

\bigskip\noindent [10] K. Soram\"aki, M. L. Bech, J. Arnold,
R.J. Glass, W.E. Beyeler, {\it Physica A} {\bf 379}, 317-333 (2007).

\bigskip\noindent [11] A.G. Haldane, R.M. May, {\it Nature} {\bf
469}, 351-355 (2011).

\bigskip\noindent [12] P. Krugman, {\it The self-organizing economy}, 
Blackwell, Oxford (1996).

\bigskip\noindent [13] S.P. Borgatti, M.G. Everett, 
{\it Soc. Net.} {\bf 21}, 375-395 (2000).

\bigskip\noindent [14] M.E.J. Newman, {\it SIAM Rev.} {\bf 45}, 167-256 (2003).

\bigskip\noindent [15] R. Olfati-Sabera, R.M. Murray, {\it IEEE 
Trans. Aut. Cont.} {\bf 49}, 1520-1533 (2004).








\vfill\eject

\noindent{\bf Supplementary material}

\vskip 1cm

\bigskip\noindent{\it Proof of theorem 1:} 

\smallskip\noindent Let 
$$G_{off} = \pmatrix{ 0 & C \cr C^\dagger & 0 \cr} \eqno(S1)$$  
be the off-diagonal part of the Hermitian gradient $G$.
Similarly, let
$$G_{on} = \pmatrix{ A & 0 \cr 0 & B \cr} \eqno(S2)$$
be the on-diagonal part of $G$.
The singular value decomposition for $C$ implies
that the eigenvalues of $G_{off}$ are either zero, 
or come in pairs $\pm \lambda_j$,
where the $\lambda_j \geq 0$ are the singular values of the matrix $C$.
The $ \pm \lambda_j$ eivenvectors of $G_{off}$ take the form 
$\vec g^\pm_j = \pmatrix{\vec u_j\cr \pm \vec v_j\cr}$, 
where $\vec u_j$ and $\vec v_j$
are the left-singular and right-singular vectors for $\lambda_j$:
$C\vec v_j = \lambda_j \vec u_j$, $C^\dagger \vec u_j
= \lambda_j \vec v_j$. 

Now look at vectors
of the form $\vec w = \pmatrix{ \alpha\vec u_j \cr \beta\vec v_j\cr}$,  
where $\alpha, \beta$ are real and non-negative.
Maximizing $\vec w^\dagger G \vec w$ over $\alpha,\beta$
yields $\vec w^\dagger G \vec w > 0$ when
$\lambda_j^2 > a_j b_j$, where $a_j = \vec u_j^\dagger A\vec u_j$
and  $b_j = \vec v_j^\dagger B\vec v_j$. 
So the system is unstable if $\lambda_j^2 > a_j b_j$ for any
$j$.  In particular, if $\sum_j a_j b_j < \sum_j \lambda_j^2$,
then the system is unstable.   

Note that $a_j$, $b_j$ are
the diagonal elements of $A$, $B$ in the bases  
$\{\vec u_j \}$ for $A$'s $n_A$-dimensional
Hilbert space, and 
$\{\vec v_j \}$ for $B$'s $n_B$-dimensional
Hilbert space.  Let $\vec \mu$ be the vector of eigenvalues
of $A$, and $\vec \nu$ be the vector of eigenvalues
of $B$.  It is straightforward to verify that
the vectors $\vec a$ with components $a_j$
and $\vec b$ with components $b_j$ are related to 
$\vec \mu$ and $\vec B$ by doubly stochastic transformations:
$\vec a = W_A \vec \mu$, $\vec b = W_B \vec \nu$.
Convexity then implies that
$|\vec a|^2 = \sum_j a_j^2 \leq |\vec \mu|^2 = \sum_j \mu_j^2
= {\rm tr} A^\dagger A$.  Similarly,
$|\vec b|^2 = \sum_j b_j^2 \leq |\vec \nu|^2 = \sum_j \nu_j^2
= {\rm tr} B^\dagger B$.  These inequalities, combined with the
Cauchy-Schwartz inequality, show that
\begin{eqnarray}
 {\rm tr} C^\dagger C &>& \sqrt{ {\rm tr} A^\dagger A}
\sqrt{ {\rm tr} B^\dagger B} \nonumber \\
\longrightarrow \sum_j \lambda_j^2 &>& |\vec \mu| |\vec \nu|
\geq |\vec a| |\vec b| \geq \vec a \cdot \vec b = \sum_j a_j b_j,\nonumber 
\quad\quad\quad\quad(S2)
\end{eqnarray}
and the system is unstable.  This proves the theorem.

Theorem 1 immediately implies a set of stability tests.
Define $a^2 = (1/n_A^2) {\rm tr} A^2 =
(1/n_A)\sqrt{ \sum_{ij=1}^{n_A} |a_{ij}|^2}$ to be
the average magnitude squared of the entries of $A$.
$a$ can be thought of as the `strength' of $A$'s stabilizing dynamics.
Similarly, $b = (1/n_B) \sqrt{{\rm tr} B^2}$
gives the `strength' of $B$'s stabilizing dynamics,
and $c = (1/\sqrt{n_An_B}) \sqrt{ {\rm tr} C^\dagger C}$ 
is the strength of the potentially destabilizing dynamics
of the interactions.
Theorem 1 is equivalent to the statement
that when $c^2 > ab$, interactions cause instability:
the system is unstable
if the strength of the destabilizing interaction dynamics
is greater than geometric mean of the strengths of the 
stabilizing local dynamics.   
Similarly, $c > (a+b)/2 \geq \sqrt{ab}$,
and $c^2 > a^2/2 + b^2/2 \geq ab$ also imply instability.

The bound of theorem 1 is tight in
the sense that for a given $C$, and fixed $a$, $b$ 
such that $ab > c^2$, the entries
of $A$ and $B$ can always be chosen to make the system
stable (bound (1c)).  The `minimal' strategy for
attaining stability is align the eigenvectors of $A$ and $B$ with the
left and right singular vectors of $C$.  Arrange $\vec \mu \propto
\vec \nu$, and $\mu_j \nu_j = \lambda_j^2 + \epsilon$.  Then
the system is stable and
$$\sqrt{ {\rm tr} A^\dagger A } 
\sqrt{ {\rm tr} B^\dagger B } 
= {\rm tr} C^\dagger C + O(\epsilon). \eqno(S3)$$  
Similarly, the tight bounds (1a), (1b) can be attained by aligning
the eigenvector of $G_{off}$ with the largest eigenvalue together with the 
eigenvectors of $A$,$B$ with the smallest eigenvalue product. 

\bigskip\noindent{\it Proof of theorem 2:}

First, look at the oscillatory behavior of the matrix
$$G' = \pmatrix{ A & \tilde C \cr  -\tilde C^\dagger & B \cr} \eqno(S4)$$
that contains the Hermitian gradients for the on-diagonal
terms and the anti-Hermitian gradient for the interaction
terms.   We will show that theorem 2 holds for this gradient.
As will be seen, adding the anti-Hermitian parts
of the on-diagonal terms and the Hermitian parts of
of the interaction terms cannot decrease the damping ratio.

Use the same analysis as in theorem 1.  Now let  
$\vec u_j$ and $\vec v_j$
be the left-singular and right-singular vectors for $\tilde C$
with singular value $\lambda_j$.
The eigenvectors of 
$G'_{off} = \pmatrix{ 0 & \tilde C \cr \tilde C^\dagger & 0 \cr}$
now take the form
$\vec g^\pm_j = \pmatrix{\vec u_j\cr \pm i \vec v_j\cr}$ with
eigenvalues $\pm i\lambda_j$.
In analogy to the proof of theorem 1, look at vectors
of the form $\vec w = \pmatrix{ \alpha\vec u_j \cr i \beta\vec v_j\cr}$,
where $\alpha, \beta$ are real.
Maximizing the damping ratio $|\vec w^\dagger G'_{off}\vec w|/| 
\vec w^\dagger G'_{on}\vec w|$ over $\alpha,\beta$ 
yields a value $ \gamma = \lambda_j/\sqrt{a_jb_j}$, 
where as before $a_j = \vec u_j^\dagger A\vec u_j$
and $b_j = \vec v_j^\dagger B\vec v_j$.  Following the proof of
theorem 1, we see that when
${\rm tr} \tilde C^\dagger \tilde C \geq
\gamma^2 \sqrt{ {\rm tr} A^2} \sqrt{ {\rm tr} B^2}$ the damping
ratio for the state $\vec w$ is equal to $\gamma$.  
This proves the theorem
for matrices of the form $G'$.   Adding the anti-Hermitian parts
of the diagonal terms can only increase the magnitude
of the numerator of the damping ratio,
while adding the Hermitian parts of the off-diagonal, interaction
terms can only decrease the minimum value of the denominator.   
Indeed, including the Hermitian part of the off-diagonal term
$G_{off}$ and performing the optimization above
yields a tighter bound: oscillation occurs with
damping ratio $> \gamma$ if
$$ {\rm tr}~ \tilde C^\dagger \tilde C 
> \gamma^2 ( 2\sqrt{{\rm tr} A^2 {\rm tr} B^2} - 
| {\vec g^+_j}{^\dagger}  G_{off} \vec g^+_j | )^2. \eqno(S5)$$
So theorem 2 holds for all gradients $\nabla g$.

Tight bounds for the oscillatory
threshold are attained in analogue to (1a), (1b), (1c) by
aligning the `strongest' and `weakest' eigenvector/eigenvalue
combinations of $A$, $B$, and $\tilde C$.

\bigskip\noindent{\it Linearized equations for the gravo-thermal
catastrophe:}

Consider a system bound together by gravitation such as a
star, nebula, or galaxy.  Such systems typically possess
a core-halo structure [7].
Let $E_A$, $T_A$ be the energy and temperature of the core,
and $C_A$ its specific heat.
Similarly, let $E_B$, $T_B$ and $C_B$ be the energy, temperature and
specific heat of the halo.  $\alpha \geq 0$ gives the linearized rate of energy
transfer between core and halo as a function of their temperature difference.  
$\eta \geq 0$ governs heat production in the core, due, e.g., to nuclear
reactions, and $\zeta \geq 0$ governs heat loss from the halo to space
beyond.  The linearized equations of motion are
$${dE_A\over dt} = C_A {dT_A\over dt} = 
\alpha(T_B - T_A) + \eta T_A, \quad
{dE_B\over dt} = C_B {dT_B\over dt} = 
\alpha(T_A - T_B) - \zeta T_B. \eqno(S4)$$
Eliminating $E_A,E_B$ then yields equation (6) of the text.

\bigskip\noindent{\it Instability of the bank transfer network:}

To derive the interaction-driven instability threshold for
the bank transfer network, divide the network into the set of 
banks with above-average numbers of links and volumes (the core,
system ${\cal A}$), and those with below average numbers of links and volumes 
(the halo, system ${\cal B}$).    Let $v_{ij}$ be the measured volume 
flow from bank $j$ to bank $i$ during a particular reporting period. 
To construct a linearized dynamics would require knowledge of 
$V_j$, the amount of funds held in bank $j$, together with the rates
$v_{jj}$ of creation and consumption of funds within the $j$'th bank.  
These numbers are not available from the data set analyzed in [10].  
Even though $V_j$ and $v_{jj}$ are unknown, however, the stability
analysis of theorems 1 and 2 can still be applied.   

In the linearized dynamics,
the gradient matrix $\nabla g$ has entries
$v_{ij}/V_j$.  The matrices $A$,$B$, $C$, and $\tilde C$
are derived from $ \nabla g$ as before: 
$A$ governs the Hermitian dynamics within
the core, $B$ governs the dynamics within the halo, and $C$,$\tilde C$ govern
flows between core and halo.  
Let $\kappa_A$ be the fraction
of non-zero terms within $A$, so that $\kappa_A n_A^2$ is the
number of links within the core.  Define 
$\kappa_B$ and $\kappa_C$ in the same way.  
Let $\bar a^2$ be the average
magnitude squared of a non-zero term in $A$: 
$\bar a^2 = a^2/\kappa_A$, where as above $a = (1/n_A) \sqrt{{\rm tr}
A^\dagger A}$ is the average
strength of all the terms in $A$, including those that
are zero.  Similarly, $\bar b^2 = b^2/\kappa_B$,
$\bar c^2 = c^2/\kappa_C$ are the average magnitude squared
of non-zero terms in $B$ and $C$.  Theorem 1 then implies that the
dynamics are unstable if 
$$ \kappa_C \bar c^2 > 
\sqrt{\kappa_A \kappa_B} \bar a \bar b.\eqno(S6)$$ 
Similarly, theorem 2 implies that the dynamics exibit oscillations
with underdamping degree $> \gamma$ if
$$ \kappa_{\tilde C} \bar {\tilde c}^2 >  \gamma^2
\sqrt{\kappa_A \kappa_B} \bar a \bar b.\eqno(S7)$$

The disassortative nature of the network [10] now sets the stage for
connectivity-driven oscillations and instability: when the strength of internal
stabilizing dynamics is insufficient, the dynamics of the network
gives rise to oscillatory and/or unstable flows between core and halo.
The disassortative nature
of the network implies that there are fewer internal links
within the core, $\kappa_A n_A^2$, and within the halo,
$\kappa_B n_B^2$, than there are between core and halo,
$\kappa_C n_A n_B$.  That is,  disassortativity implies
$\kappa_C^2 > \kappa_A \kappa_B$.   
Equations (S6) and (S7) then imply theorem 3. 
In a core-halo system, stability
requires that the strength of links within the core
be significantly higher than the strength of links
between core and halo, a feature observed in the
actual bank transfer network (3/4 of the volume of transfers occurs
within the core).

If the system is marginally stable, so that
$\kappa_C \tilde c^2 \approx 
\sqrt{\kappa_A \kappa_B} \tilde a \tilde b,$
then any dip in the connectivity or the strength of connections 
within the core will drive it unstable.

\vfill\eject

\begin{figure}
\begin{center}
\includegraphics[width=250pt]{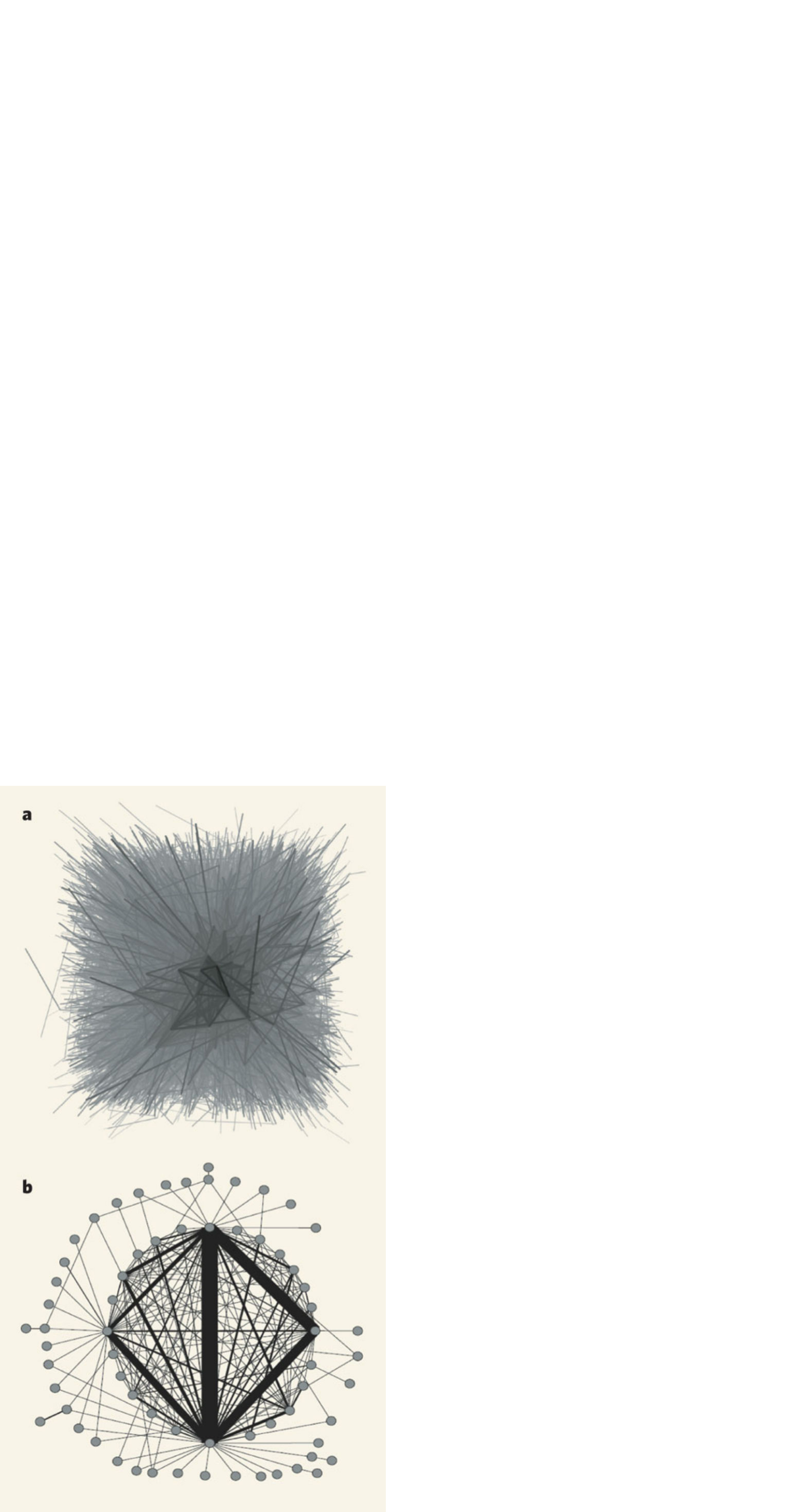}
\caption{The interbank transfer network in 2007
The US interbank transfer network 
consists of over $6600$ financial institutions making over
$70,000$ daily transfers.   As mapped here from pre-financial crisis
data, the network exhibits a pronounced
core-halo structure (2a: reprinted with permission from reference [10]).  
The core of $66$ institutions (2b) includes
an inner core of $25$ fully connected institutions, and accounts
for $3/4$ of the transfer volume.  As with the gravo-thermal 
catastrophe, a slowdown of the core can drive the entire system
unstable, leading to the financial analogue of gravitational 
collapse (`the black hole of finance').} 
\label{fig:fig1}
\end{center}
\end{figure}

\end{document}